\documentclass[aps,pre,reprint,superscriptaddress,showpacs,preprintnumbers,amsmath,amssymb,nofootinbib]{revtex4-2}
\usepackage[dvipdfmx]{graphicx}
\usepackage{dcolumn}   % needed for some tables
\usepackage{bm}        % for math
\usepackage{amssymb, amsmath, amsfonts}   % for math
\usepackage{lipsum}        % for math
\usepackage{color}
\usepackage{appendix}
\DeclareGraphicsExtensions{{.pdf}}
\begin{document} 
% Include your paper's title here
\title{Deep generative model super-resolves spatially correlated multiregional climate data} 
\author
{Norihiro Oyama}
\email{Norihiro.Oyama.vb@mosk.tytlabs.co.jp}
\affiliation{Toyota Central R\&D Labs, Inc., Bunkyo-ku, Tokyo 112-0004, Japan}
\author{Noriko N. Ishizaki}
\affiliation{Center for Climate Change Adaptation, National Institute for Environmental Studies, Tsukuba 305-8506, Japan}
\author{Satoshi Koide}
\affiliation{Toyota Central R\&D Labs, Inc., Bunkyo-ku, Tokyo 112-0004, Japan}

\author{Hiroaki Yoshida}
\affiliation{Toyota Central R\&D Labs, Inc., Bunkyo-ku, Tokyo 112-0004, Japan}

% Include the date command, but leave its argument blank.

\date{\today}

%\keywords{Keyword1, Keyword2, Keyword3}

\begin{abstract}
Super-resolving the coarse outputs of global climate simulations, termed downscaling, is crucial in making political and social decisions on systems requiring long-term climate change projections. Existing fast super-resolution techniques, however, have yet to preserve the spatially correlated nature of climatological data,
which is particularly important when we address systems with spatial expanse, such as the development of transportation infrastructure.
Herein, we show an adversarial network-based machine learning enables us to correctly reconstruct the inter-regional spatial correlations in downscaling with high magnification of up to fifty while maintaining pixel-wise statistical consistency.
Direct comparison with the measured meteorological data of temperature and precipitation distributions reveals that integrating climatologically important physical information improves the downscaling performance, which prompts us to call this approach $\pi$SRGAN (Physics Informed Super-Resolution Generative Adversarial Network).
The proposed method has a potential application to the inter-regionally consistent assessment of the climate change impact. 
{Additionally, we present the outcomes of another variant of the deep generative model-based downscaling approach in which the low-resolution precipitation field is substituted with the pressure field, referred to as $\psi$SRGAN (Precipitation Source Inaccessible SRGAN). 
Remarkably, this method demonstrates unexpectedly good downscaling performance for the precipitation field.}
\end{abstract}
\maketitle

\section*{Introduction}
The increase of greenhouse gases in the air composition due to human activities is now believed to have led to the rise in the frequency of unusual disasters~\cite{Disaster1, Disaster2, Disaster3, Disaster4}.
To prevent an irreversible collapse of the current ecosystem and resulting impoverishment of human lives, many countries have set specific medium- and long-term goals for the reduction of greenhouse gas emissions, and similar paradigm shifts in decision making have occurred even at the private sector level. 

Numerical approaches are regarded as the most powerful and reliable scientific option at the moment in quantitatively evaluating the efficacy of political or management plans that aim to tackle climatological issues.
The Global Climate Model (GCM) is the prime example, which has accurately reproduced past and current climate changes, and its reliability of quantitative future estimates is sufficiently high \cite{WGI}.
Such future projections with high accuracies rely on the overall consideration of the global atmospheric and oceanic circulation (and even still more complicated ingredients such as {chemical~\cite{chaser} and biological~\cite{SEIB} processes})~\cite{MIROC5,MRI-CGCM3,GISS-E2,HadCM3,CMCC-CM}, and thus, the horizontal spatial resolution is sacrificed by the required computational costs; the typical resolution of the GCMs is only down to the order of one degree in longitude and latitude, corresponding to a grid size of more than a hundred kilometers on the equator.
Therefore, to exploit the GCM outputs to assess the impact of climate change and to make proper decisions, 
it is obviously vital to super-resolve the coarse grid spacing of simulations and to reach the fine resolution of interest. 
Here, special attention should be given to reproducing the inherent spatial correlation of the meteorological variables, as well as the local statistics, in decision making by integrating multiregional information~\cite{MR1,MR2,MR3,MR4,MR5}, such as transportation infrastructure development and sustainable energy networks, future urbanization, and agricultural intensification.

\begin{figure*}[t]
    \centering
    \includegraphics[width=\textwidth,bb=0 0 1489 965]{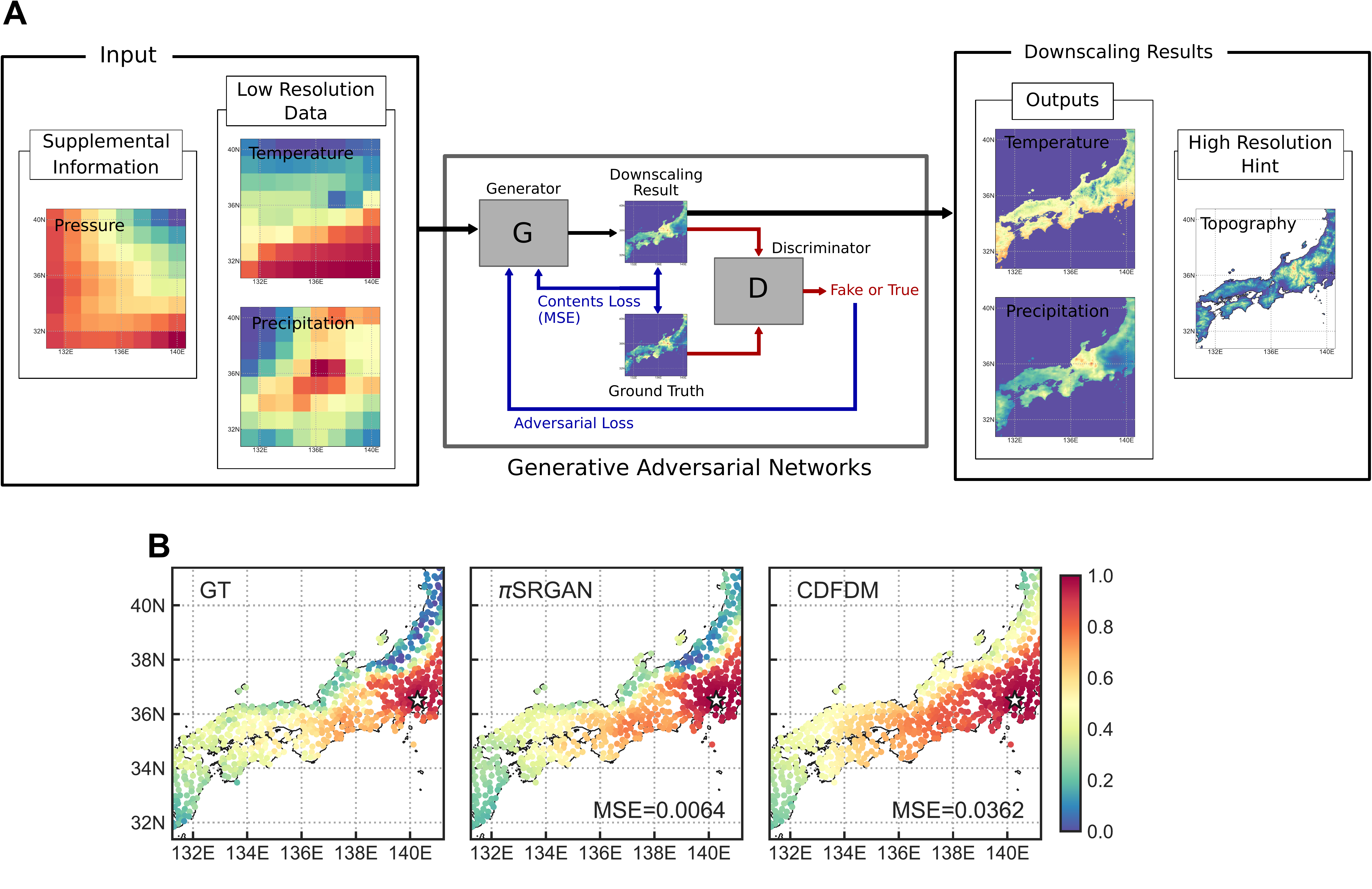}
    \caption{
    Schematic diagram of $\pi$SRGAN and the distribution of spatial correlation coefficients.
        (A) High-resolution topography and low-resolution sea level pressure, in addition to the low resolution data corresponding to the output, are supplied to the generative adversarial networks.
        (B) The reconstructed distributions of spatial correlation coefficients, indicating the correlation strength from the reference site at Tokyo {[35.735$^\circ$N, 139.6683$^\circ$E]}, obtained with $\pi$SRGAN and a conventional CDFDM  are compared to the ground truth (GT).
    }
    \label{fig:this_method}
\end{figure*}

A variety of techniques to super-resolve GCM outcomes,
which are referred to as the \emph{downscaling} (DS) methods in meteorology and climatology, have been developed~\cite{DD1,DD2,DD3,DD4,SD1,Wilby2004,SD3,SD4}.
They are categorized roughly into two groups: dynamical~\cite{DD1,DD2,DD3,DD4} and statistical DS methods~\cite{SD1,Wilby2004,SD3,SD4}.
The dynamical downscaling method is based on physical footings: several coupled differential equations are numerically integrated with the results of the GCM (or any other crude-resolution simulation results) being used as the boundary conditions.
However, the computational cost again creates a trade-off between the accuracy and the feasibility. 
In contrast, 
in the statistical approaches, we turn a blind eye to the physical laws behind the data. Instead, empirical links between the large- and local-scale climates are identified and applied to the crude-resolution climate model outputs. 
Since the systematic errors of the naively interpolated GCM output (referred to as the bias) are locally corrected such that the statistical properties are precisely reproduced, the spatial correlation, i.e., the information on the events occurring at distant places, is discarded~\cite{SpatialCorrelation1,SpatialCorrelation2,SpatialCorrelation3}.
The statistical downscaling methods overcoming the latter problem remain to be developed.

In this paper, we propose a machine learning approach that super-resolves the GCM outputs and reproduces both the local statistics and the instantaneous spatial correlations between distant regions.
{Among several options for improving the resolution of geophysical or climatological data\cite{kaur1,kaur2,ex1,ex2},} our method is based on the generative adversarial network (GAN) approach, which has been proven to be a very powerful downscaling tool through several previous studies~\cite{PNAS_GAN_DS,DeepDT}.
To accurately reproduce the physical nature, we use auxiliary but climatologically important data, sea-level pressure distribution and topographic information, in addition to the target variables, temperature and precipitation distributions (see Fig.~\ref{fig:this_method}A and the next section for more details).
Since this method falls within the criteria of the first-level physics informed super-resolution methods \cite{Onishi2019}, we name our method $\pi$SRGAN (Physics Informed Super-Resolution Generative Adversarial Network).
The direct comparison with the measured meteorological data shows that the local statistical properties are obtained using the practical output from the GCM simulations as accurately as the conventional statistical downscaling method that is focused on matching these properties.
We then highlight that the spatial correlation of variables is accurately reproduced, which could not be achieved with conventional downscaling methods (see Fig.~\ref{fig:this_method}B).
The present method is therefore the next generation downscaling method that has a potential application in climate change assessment considering both local-scale and interregional events.
{We also considered another variant of the SRGAN that projects the high-resolution temperature and precipitation field from the low-resolution information about only temperature and pressure (we call this variant the $\psi$SRGAN: Precipitation-Source-Inaccessible SRGAN).
With this special variant, we demonstrate the surprisingly robust ability of the SRGAN-based methods to express natural results.}

%------------------------------------------------------------------------------------------
%%% Table: Summary of protocols considered
%------------------------------------------------------------------------------------------
\tabcolsep = 5pt
\begingroup
\renewcommand{\arraystretch}{1.5}
\begin{table*}[tbh]
  \textbf{\caption{\label{table:methods}Summary of protocols compared in this study}}
  \centering
  \begin{tabular}{cccc}
    \hline
     Abbr. name & Explanation & Low-resolution data& High-resolution data\\
    \hline 
    GT &Ground truth (observation results offered by AMGSD~\cite{AMGSD}) &- &-\\
    LR &Low resolution data of JRA-55~\cite{jra55} &- &-\\
    SRGAN &Standard SRGAN-based method~\cite{PNAS_GAN_DS} & PRC, TMP & PRC, TMP\\
    {$\pi$SRGAN} &{Physics-Informed SRGAN}&PRC, TMP, SLP & PRC, TMP, TOPO$^\dagger$\\
    {$\psi$SRGAN} &{Precipitation-Source-Inaccessible SRGAN}&TMP, SLP & PRC, TMP\\
    CDFDM &Cumulative distribution function-based downscaling method~\cite{Iizumi2010}&PRC, TMP &PRC, TMP\\
    \hline
    \multicolumn{4}{r}{(PRC: precipitation, TMP: temperature, SLP: sea level pressure, TOPO: topography)}\\
    \multicolumn{4}{r}{$\dagger$: In $\pi$SRGAN, TOPO is used as a hint, not targeted}
  \end{tabular}
\end{table*}
\endgroup

\section*{Results}
\subsection*{{Super-Resolution Generative Adversarial Networks with various data}}
%%%%%%%%%%%%%%%%%%%%%%%%%%%%%%%%%%%%%%%%%%%%%%%%%%%%%%%%%%%%%%%%%%%%%

We employ a super-resolution method based on generative adversarial networks (Super-Resolution Generative Adversarial Networks: SRGAN, see Methods section for details) as the basic machine learning algorithm, which was proven to have potential in DS with a scale factor up to fifty~\cite{PNAS_GAN_DS}. 
Although the original SRGAN was able to restore physical consistency in the turbulent wind velocity field, which was shown in terms of the well-known Kolmogorov $5/3$ power-law~\cite{Frisch1995},
it was also reported that it showed a worse performance in reproducing the basic statistics, such as the pixelwise consistency like mean squared error, than a less sophisticated deep learning approach~\cite{PNAS_GAN_DS}.
{In this work, considering two distinct variants in addition to the standard SRGAN, we show that the integration of the low-resolution input with auxiliary information enables to overcome the drawback of relatively poor reproducibility of simple statistical properties and that the ability of SRGAN-based methods to downscale in a ``physically natural'' manner is quite robust against the change in the input low-resolution information.}

{There are a vast variety of LR information, as seen in several similar recent attempts~\cite{Vandal2017,Vandal2018,DeepDT,Yasuda2021}.
Among them, we employed the sea-level pressure, one of the fundamental hydrodynamic (or aerodynamic) variables on which the various quantities of sub-models of GCMs are based, as a piece of key auxiliary information.
Also, this variable is described with fewer assumptions in the models than other meteorological variables such as humidity.
In the literature, strong links between synoptic-scale horizontal circulation and vertical motion are discussed in terms of the sea-level pressure~\cite{Pressure4, pressure1, pressure2, pressure3}.}
{ 
In the first variant, we incorporate the low-resolution pressure field as an auxiliary physical information (Fig.~\ref{fig:this_method}A), which serves as guidance for the DS of the target variables, namely temperature and precipitation.
In this method, moreover, we introduced the topographic information as another auxiliary information since it can be utilized in a high-resolution format only if we assume it is identical over the time window of interest (order of tens to a hundred years).
The topographic information is indirectly supplied as a part of teacher data during the training by adding to one of the output channels.
In this way, we can provide both low-resolution and high-resolution auxiliary data in an unambiguous manner without any artificial operation (like resolution matching by interpolation or pooling).
Since the use of supplemental physical information during learning is regarded as primary-level physics-informed machine learning~\cite{Onishi2019}, we call this method the Physics-Informed SRGAN ($\pi$SRGAN for short).}

{The second variant of SRGAN is designed to generate high-resolution temperature and precipitation fields using solely low-resolution data pertaining to the temperature and pressure fields. 
This variant is referred to as the Precipitation-Source-Inaccessible SRGAN ($\psi$SRGAN) and demonstrates the surprisingly robust capability of SRGAN-based methods to describe ``physically natural'' precipitation fields.}

{The performances of three variants of SRGAN (standard SRGAN, $\psi$SRGAN, and $\pi$SRGAN) are evaluated via direct comparisons among them and with a non-machine learning-based method: we summarize these methods in Table~\ref{table:methods}.
The cumulative distribution function-based downscaling method
(CDFDM) is the widely used conventional statistical DS method (see the method section for the details), and the SRGAN refers to the original SRGAN-based method presented in Ref~\cite{PNAS_GAN_DS}.}

\subsection*{Data sets}
We use the climate model simulation outputs for the low-resolution input and the real observation data for the high-resolution ground truths in the case studies.
As the low-resolution data, we used the Japanese 55-year reanalysis (JRA-55) data from 1980 to 2018~\cite{jra55} with data assimilation. The grid spacing is $1.25$ degrees. 
The daily data corresponding to the reference data (in Japanese local time) were created from 3-hourly simulation data.
Specifically, data at 0Z, 3Z, 6Z, 9Z and 12Z on the target date and data at 15Z, 18Z and 21Z on the previous day of the target date were averaged to obtain the daily data in JST.
The reference high-resolution data were the Agro-Meteorological Grid Square Data (AMGSD)~\cite{AMGSD}.
The 1~km-meshed daily data over Japan are constructed using the in-situ observation network system of the Japan Meteorological Agency, which covers the entire land area over Japan from $122$ to $146$ degrees east and from $24$ to $46$ degrees north. 
{Upon being fed into the networks, all of the data undergo a process of normalization and concatenation. For further information regarding the technical aspects of these procedures, please refer to the SI Appendix.}
%------------------------------------------------------------------------------------------
%%% Table: Split of data
%------------------------------------------------------------------------------------------
\tabcolsep = 5pt
\begingroup
\renewcommand{\arraystretch}{1.5}
\begin{table*}[t]
  \textbf{\caption{\label{table:data_split}Year span for each data set}}
  \centering
  \begin{tabular}{ccc}
    \hline
      Training data & Validation data  & Test data \\
    \hline 
    1980-1997 & 1998-2000 & 2001-2018\\
    \hline
  \end{tabular}
\end{table*}
\endgroup

%%%%%%%
{We use the data from 1980 to 2018 ($14245$ days in total).
These data are split into training, validation, and test datasets in a time-series manner as summarized in Table~\ref{table:data_split} for both low-resolution (JRA-55) and high-resolution (AMGSD) data.
We emphasize that this time-series partitioning, characterized by a substantial volume of test data, represents a challenging task for downscaling mid-term future projections, and consequently, necessitates the incorporation of the climate change trend.}
The AMGSD data were adjusted such that the grid spacing was $0.025$ degrees/grid both in latitude and longitude.
We extracted the data for the region from $130.625$ to $140.625$ degrees east and from $30.625$ to $40.625$ degrees north, which results in a $400\times 400$ pixels square.
The JRA-55 data of the corresponding region are $8\times 8$ pixel squares, and thus, the scale factor for the DS tasks is $50$.

\begin{figure*}[t]
    \centering
    \includegraphics[width=0.97\linewidth,bb=0 0 806 346]{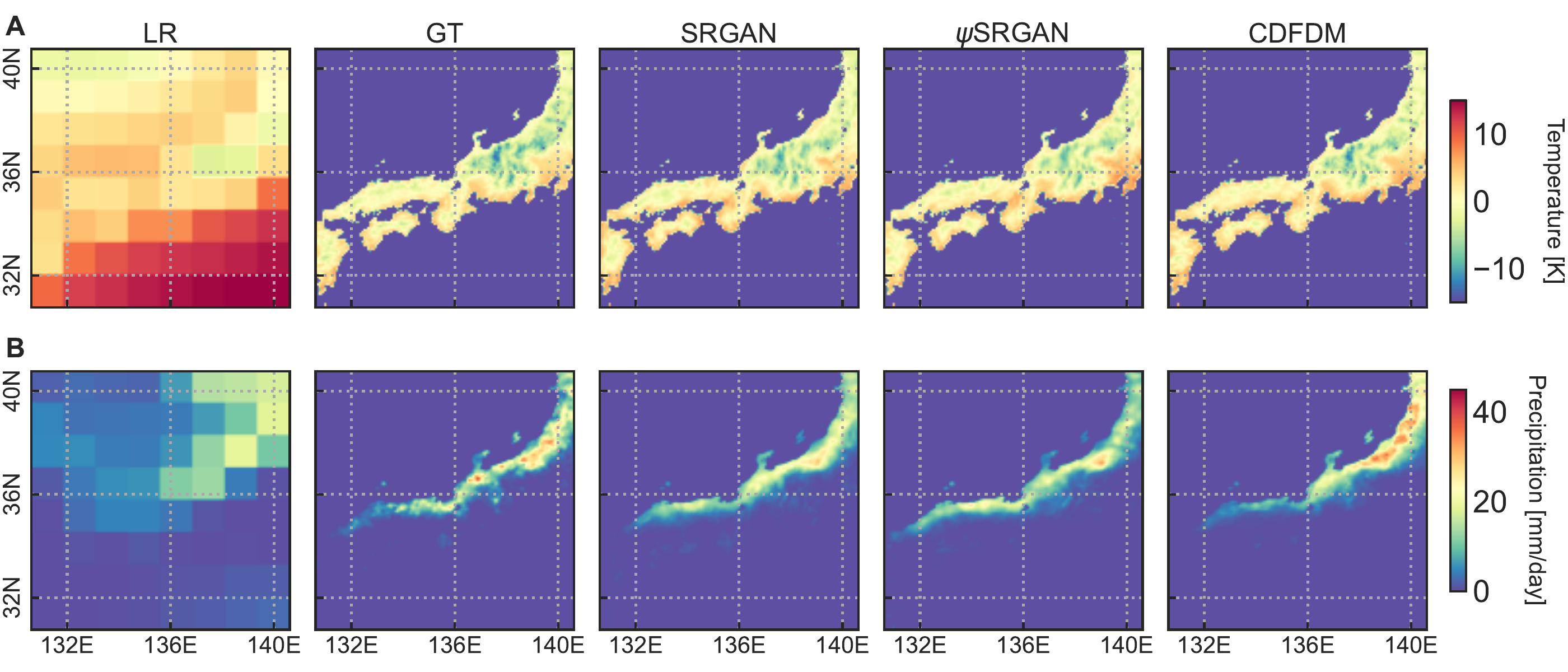}
    \caption{
    Downscaling results.
    Distributions of temperature (A) and precipitation (B) obtained with SRGAN, $\psi$SRGAN, and CDFDM are compared with the corresponding low-resolution ($8\times 8$) inputs (LR) and the high-resolution ($400\times 400$) ground truth (GT). 
    The resolution of the downscaled images is the same as that of the GT.
    The data from January 24, 2008, are displayed.
    }
    \label{fig:qualitative}
\end{figure*}
%%%%%%%

\subsection*{Qualitative visualization}\label{sec:visualization}
We first present typical qualitative visualizations for the temperature and the precipitation fields of one day in Fig.~\ref{fig:qualitative},
which highlights the ambitious downscaling with the present large scaling factor of fifty.
Here, the high-resolution information of $2500$ pixels is extracted from one single pixel in the low-resolution counterpart.
We compare the results of different protocols (summarized in Table~\ref{table:methods}), along with the visualization of the original low-resolution JRA-55 and the high-resolution AMGSD data.

The difference in the downscaled temperature from the ground truth is not very large (the upper row of Fig.~\ref{fig:qualitative}), and it is difficult to find any superiority or inferiority in performance from these qualitative plots. 
In contrast, the results for precipitation demonstrate rich information on the features of DS protocols (the lower row of Fig.~\ref{fig:qualitative}). 
The CDFDM result shows an overly smoothed profile compared to the GT: high precipitation values (represented by red colors) are observed in a vaster area.
On the other hand, SRGAN family finely reproduce the localized nature of the high precipitation areas, which the CDFDM fails to describe. 
{Remarkably, even $\psi$SRGAN also succeeded in reproducing the localized heavy rain event, although, in this method, the low-resolution precipitation field is not supplied as an input.
The GAN-based methods~~\cite{MR1,MR2,MR3,MR4,MR5} are recognized to be advantageous in reproducing such fine structures.}
The maximum precipitation values of the DS results are all very close to that of the GT.
{Please refer to Fig.~S2 in the SI Appendix for the graphical depictions of the differences between the GT and DS outcomes, which offer a more direct and intuitive insight into the distinctions among the performances of different methods.
We note that although the results for $\pi$SRGAN were excluded from Fig.~\ref{fig:qualitative} due to their substantial similarity with those for SRGAN and space limitations, they are included in Figure S2 of the SI Appendix.}

\subsection*{Single-site statistics}\label{sec:PDFs}
%%%%%%%
Here and in the following subsections, we discuss the statistical features of downscaling results, focusing on the precipitation $p$, which is generally considered to be difficult to downscale accurately.
In particular, we carefully examine the statistical consistency with the ground truth, which is crucial in actual usage of the DS results, e.g., in impact assessment of climate change in the future.
{Although the results presented in the main text are climatologically oriented indicators and not standard measures used in the field of image processing, we provide the values of pixel-wise mean squared error and corresponding peak signal-to-noise ratio in the SI Appendix.}

\begin{figure*}[!t]
    \centering
    \includegraphics[width=0.97\linewidth,bb=0 0 861 770]{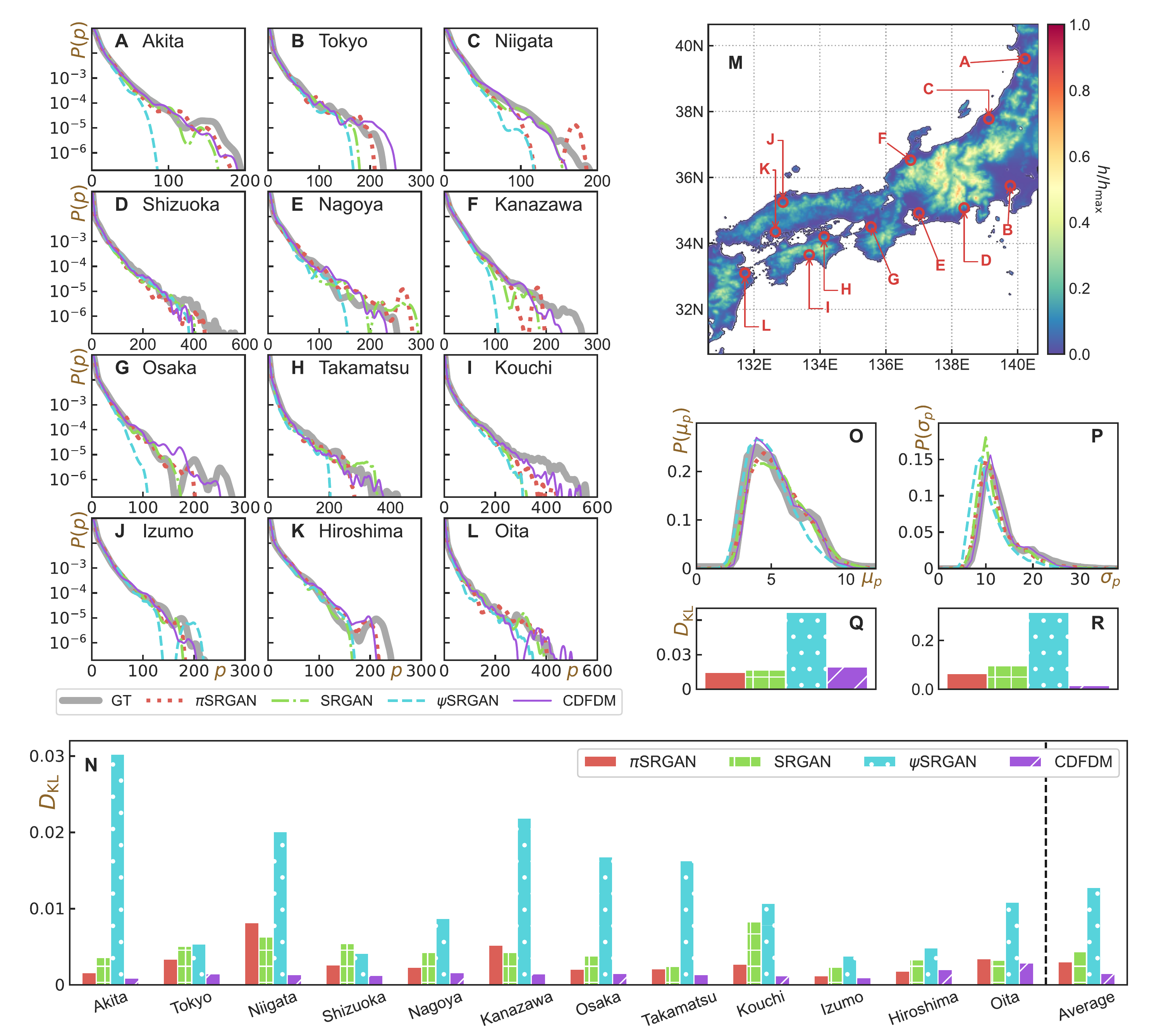}
    \caption{
    Statistics of precipitation.
   (A-L) The probability distribution functions (PDFs) $P(p)$ as a function of the precipitation $p$ at each site. 12 representative sites are chosen from all over Japan.
   The different ranges of the abscissa reflect the regional characteristics. See SI Appendix for the technical details in processing the PDFs.
   (M) The normalized topographic information and the locations of $12$ sites of panels (A-L).
    (N) Bar plot of the values of the Kullback-Leibler divergence of DS methods for $P(p)$.
   (O,P) The PDFs of the mean $\mu_p$ and standard deviation $\sigma_p$ of the precipitation over all the test data at each site; here the values of $\mu_p$ and $\sigma_p$ at different sites serve as samples of the PDFs.
   (Q, R)Bar plot of the values of the Kullback-Leibler divergence for $P(\mu_p)$ and $P(\sigma_p)$.
    }
\label{fig:test_precipitation_area}
\end{figure*}

We first measure the probability distribution functions (PDFs) of the precipitation data at $12$ representative sites, $P_{\cal S}(p)$.
Here, the PDFs are calculated using the set $\lbrace p_k(l)|l\in{\cal S}\ {\rm and}\ k\in{\cal D}_{\rm test}\rbrace$, where ${\cal S}$ stands for the site of interest (each site includes 100 grid points: see Table~S3 in the SI Appendix), 
${\cal D}_{\rm test}$ is the set of dates that are used for the test data (year span of 2001-2018; Table~\ref{table:data_split}), and $p_k(l)$ is the value of the precipitation at the pixel $l$ and for the date $k$ (we omit the subscript unless necessary below).
The results are shown in Fig.~\ref{fig:test_precipitation_area} (A-L). 
The $12$ sites in Fig.~\ref{fig:test_precipitation_area} are chosen from the seaside areas within the system boundary of this study, as depicted in Fig.~\ref{fig:test_precipitation_area}(M).
Table S3 in the SI Appendix provides more precise information (latitude, longitude, etc.) about these sites.

Overall, Fig.~\ref{fig:test_precipitation_area}(A-L) shows that all methods express the regional dependence.
Regarding each method, the CDFDM provides results matching the GT very well, including the heavy rainfall regime where $p>50\, \sim$mm/day up to the values at which $P^{\rm GT}(p)$ {becomes around $10^{-4}$}.
This is expected because in the CDFDM the data are processed such that the resulting PDFs become completely consistent with the training data.
{If we shift our attention to the results of SRGAN family, we first notice that SRGAN and $\pi$SRGAN are as accurate as the CDFDM for most sites and most values of $p$.
Moreover, surprisingly, even $\psi$SRGAN succeeded in the projection of precipitation in the range $P^{\rm GT}(p)>10^{-3}$ at most sites although it was not provided with any direct information about the precipitation.
In particular, we would like to stress that an extremely high accuracy has been successfully obtained for Shizuoka, a representative site on the Pacific Ocean side (south side), where pressure-dominated summer-type precipitation events occur frequently. This indicates that the pressure field effectively serves as crucial information for the precipitation projection, such as the location of the typhoons. On the other hand, the accuracy is significantly lower at sites on the Sea of Japan side (north side), Akita, Niigata, and Kanazawa (A, C, F), which are less directly affected by typhoons. These trends are interestingly consistent with our knowledge, and it appears as if SRGAN is extracting physical laws from the data and making predictions, just as humans do. 
Then it is natural that this success of projection of the high-resolution precipitation from the low-resolution pressure drove us to believe the integration of the input information employed in $\pi$SRGAN would further improve the downscaling performance of SRGAN.
However, since all SRGAN, $\pi$SRGAN, and CDFDM offer highly accurate results, it is difficult to visually judge from the graphs which one is better than the others: we make a quantitative comparison in the next paragraph.
Before moving forward to the quantitative analysis,we remark on the discrepancies observed for tails in the large precipitation (small probability of $P^{\rm GT}(p)<10^{-4}$) regime even in the cases of the CDFDM. 
These rare events corresponding to disaster-level torrential rains are very important from the perspective of disaster prevention but are beyond the limit of the current statistical DS methods, on which we provide an overview in the Discussion section.}

To investigate the difference in the performance of $\pi$SRGAN and SRGAN, we quantify the accuracy of each method using the Kullback--Leibler divergence $D_{\rm KL}$:
\begin{equation}
    D_{\rm KL}(P^{\rm GT}||P^{DS})\equiv \int dp P^{\rm GT}(p)\log\frac{P^{\rm GT}(p)}{P^{DS}(p)},
\end{equation}
where $P^{\rm GT}(p)$ is the PDF of the GT and $P^{DS}(p)$ is that calculated using the downscaling results ($DS\in\lbrace \pi$SRGAN, SRGAN, $\psi$SRGAN, CDFDM$\rbrace$).
Generally, the more different $P^{\rm GT}(p)$ and $P^{DS}(p)$ are, the larger $D_{\rm KL}$ becomes; $D_{\rm KL}$ vanishes when the two PDFs are exactly identical.
Since the difference between two PDFs, $P^{\rm GT}(p)$ and $P^{DS}(p)$, is weighted by the ground truth distribution, the KL divergence places more importance on the frequently occurring events than on rare events.
Technical details such as the data preprocessing employed are provided in SI Appendix.
The KL divergence between the GT and DS results using distinct methods are shown by bar plots in Fig.~\ref{fig:test_precipitation_area}(N) and summarized in Table~\ref{table:summary}, where the values averaged over the $12$ sites are presented.
The precise values of $D_{\rm KL}$ for each single site are provided in Table S4 in the SI Appendix. 
{As expected from the fact that the CDFDM concentrates on matching these statistics for the training data, it gives the best values for most cases.
However, it should be noted that, at Hiroshima (denoted by K), $\pi$SRGAN marks a better score than CDFDM.
This result evidences the remarkable performance of $\pi$SRGAN concerning the basic statistical characteristics that the standard SRGAN can handle relatively inadequately.
Indeed, among SRGAN family, $\pi$SRGAN marks the best performance if we compare them by the average value over 12 sites: $\bar{D}_{\rm KL}(P^{\rm GT}||P^{\pi{\rm SRGAN}})$ is smaller than $\bar{D}_{\rm KL}(P^{\rm GT}||P^{{\rm SRGAN}})$ by approximately {40\%} (the bars signify that the presented values represent the mean across 12 sites.). 
However, $\pi$SRGAN is not always better than SRGAN and it shows worse results than SRGAN at Niigata, Kanazawa, and Oita (C, F, L).
It is noteworthy that these particular locations are precisely where the performance of $\psi$SRGAN is significantly lacking. 
This observation suggests that the inclusion of low-resolution pressure fields may have led to undesired effects.
We also note that, on the other hand, $\psi$SRGAN exhibits a lower value of $D_{\rm KL}$ than that of the standard SRGAN at Shizuoka (Fig.~\ref{fig:test_precipitation_area}(D)) where the pressure field is expected to play a crucial role in the determination of rainfall events.
These findings about the effects of the introduction of auxiliary fields should be utilized for the future refinement of the method.
To give a conclusion for this section, remarkably, even the standard SRGAN shows the same order of values of $D_{\rm KL}$ as those of CDFDM. 
Moreover, the provision of climatologically important auxiliary information can further improve the precision by $40\%$, evidenced by the results of $\pi$SRGAN.}

%------------------------------------------------------------------------------------------
%%% Table: Values of KL Divergence
%------------------------------------------------------------------------------------------
\tabcolsep = 5pt
\begingroup
\renewcommand{\arraystretch}{1.5}
\begin{table*}[t]
  \textbf{\caption{\label{table:summary}Average KL divergence of PDFs}}
  \centering
  \begin{tabular}{c|cccc}
    \hline
     & $\pi$SRGAN & SRGAN  & $\psi$SRGAN & CDFDM \\
    \hline 
    $\bar{D}_{\rm KL}$ &$3.06\times 10^{-3}$ &${4.38\times10^{-3}}$ &${1.28\times10^{-2}}$ &${\bf 1.50\times10^{-3}}$\\
    \hline
  \end{tabular}
\end{table*}
\endgroup

\subsection*{Statistics over all sites}
As another meteorologically important statistical point of view, we further measure the statistics over all sites: the PDFs of the mean $\mu_p$ and the standard deviation $\sigma_p$ of the precipitation calculated over all test data on each pixel $l$:
\begin{align}
    \mu_p(l)&\equiv\frac{1}{N_{\rm test}}\sum_k^{N_{\rm test}}p_k(l),\\
    \sigma_p(l)&\equiv \sqrt{\frac{1}{N_{\rm test}}\sum_k^{N_{\rm test}}\left(p_k(l)-\mu_p(l)\right)^2},
\end{align}
where $k\in {\cal D}_{\rm test}$ is again the sample index, and $N_{\rm test}$ is the number of samples in ${\cal D}_{\rm test}$.
The probability distribution of $\mu_p$ and $\sigma_p$, denoted by 
$P(\mu_p)$ and $P(\sigma_p)$, are shown in Fig.~\ref{fig:test_precipitation_area}(O,P).
Note that here the values calculated on each pixel serve as samples for these PDFs.
{The corresponding KL divergence $D_{\rm KL}$ for $P(\mu_p)$ and $P(\sigma_p)$ are presented in Fig.~\ref{fig:test_precipitation_area}(Q,R) as well.

{Remarkably, regarding the statistics of pixelwise average over all dates in the test dataset $P(\mu_p)$, $\pi$SRGAN (and moreover, SRGAN as well) achieves a better score than CDFDM.
{However, on the other hand, regarding $P(\sigma_p)$, CDFDM is the best and it shows almost identical results as GT. The small shifts of the whole curve of $P(\sigma_p)$ to the left} of SRGAN-based methods are consequences of the underestimation of the high-precipitation events shown in Fig.~\ref{fig:test_precipitation_area}(A-L).}
These results suggest that SRGAN-based methods exhibit a bias towards typical values in downscaling results, as opposed to presenting bold projections of extreme events, compared to CDFDM.
This is actually an anticipated tendency considering the design of the standard training scheme employed in machine learning-based methods.}

\subsection*{Spatial correlation}\label{sec:spatial}
Next, we examine in detail the spatial correlation of the downscaled results. 
The importance of the spatial correlation of the meteorological variables, i.e., the relation between two distant sites, has been realized very recently~\cite{MR1,MR2,MR3,MR4,MR5}, e.g., in the context of impact assessment of climate change.
However, conventional DS methods such as CDFDM have proven to overestimate the correlation even though the statistical consistency with the GT is maintained~{\cite{SpatialCorrelation1,SpatialCorrelation2,SpatialCorrelation3}}.
Such a tendency is actually seen in the qualitative visualizations in Fig.~\ref{fig:qualitative}, where the overly smoothed profiles are obtained.
We thus systematically evaluate the accuracy in expressing the spatial correlation of the precipitation by measuring the Pearson's correlation coefficients of the precipitation $C_M^R(l,l^\prime)$ between two sites, $l$ and $l^\prime$, which is defined as:
\begin{equation}
    C^R_{M}(l,l^\prime)=\left< \frac{\frac{1}{N_M}\sum_k^{N_M} (\delta p^R_k(l)\delta p^R_k(l^\prime))}{\sqrt{\frac{1}{N_M}\sum_k^{N_M} (\delta p^R_k(l)
    )^2}\sqrt{\frac{1}{N_M}\sum_k^{N_M} (\delta p^R_k(l^\prime))^2}}\right>_M
\end{equation}
where $\delta p_k^{R}(l)\equiv p_k^{R}(l)-\bar{p}_M^{R}(l)$ is the deviation of the $k$-th sample at site $l$ from its reference average value $\bar{p}_M^R(l)$.
The subscript $M$ indicates that the average is taken over the data of month $M$, the superscript $R\in\lbrace {\rm GT},\pi{\rm SRGAN},{\rm SRGAN},\psi{\rm SRGAN},{\rm CDFDM}\rbrace$ distinguishes the datasets and $N_M$ represents the total number of test data samples belonging to month $M$.
Since the distribution of the correlation coefficients is known to have features specific to each month, we measure the monthly values of the coefficients.
Below, we focus on the results for $M={\rm January}$, for which a previous work has pointed out the existence of a distinguished spatial pattern of precipitation correlation~\cite{Ishizaki2023}.

\begin{figure*}[t]
    \centering
    \includegraphics[width=0.97\linewidth,bb=0 0 720 714]{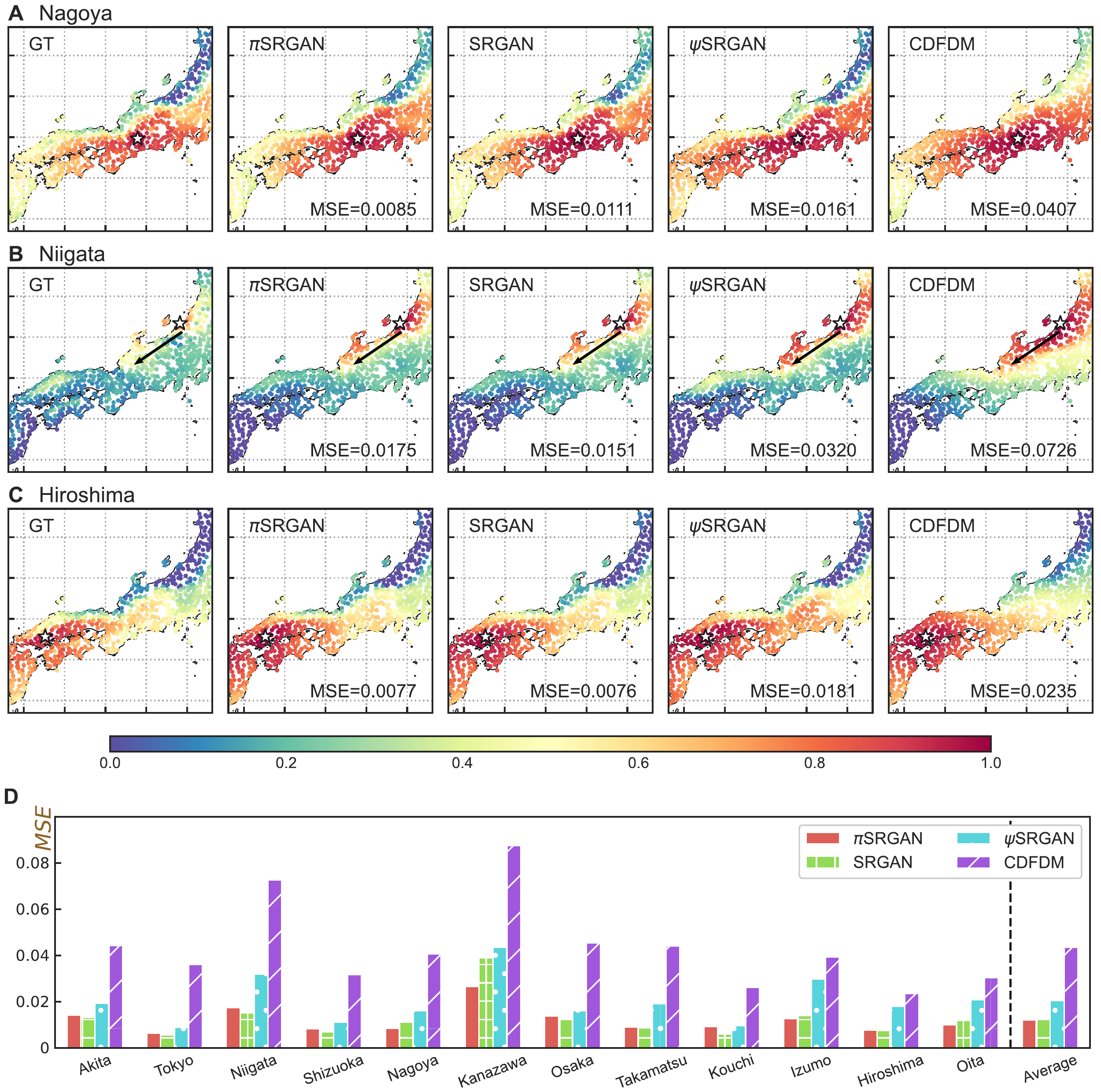}
    \caption{
    Spatial distribution of the correlation coefficients for precipitation.
    The distributions of January obtained with the $\pi$SRGAN, SRGAN, $\psi$SRGAN, and CDFDM are compared against the ground truth (GT) in the case of the reference point of correlation at Nagoya {[35.1667$^\circ$N, 136.965$^\circ$E]} (A), Niigata {[37.9133$^\circ$N, 139.0483$^\circ$E]} (B), and Hiroshima {[34.365$^\circ$N, 132.4333$^\circ$E]} (C).
    The dot color indicates the values of $C_{\rm Jan}^R(l,l^\prime)$ between the location of the dots and the reference site.
    The reference points are represented by star symbols.
    (D) The mean square error (MSE) of the correlation coefficients of the downscaled precipitations from those of the ground truth.
    Although in panels (A-C), the longitude and latitude values are omitted for reasons of space, they correspond to the same values as in Figure 3M.
 }
    \label{fig:spatial_distribution}
\end{figure*}

{
Figures~\ref{fig:spatial_distribution}(A-C) show the spatial distribution of the correlation coefficients $C_{\rm Jan}^{R}(l,l^\prime)$, with Nagoya, Niigata, and Hiroshima being the reference points $l$ (the locations of the reference points are marked by the star symbols).
The correlations measured for the CDFDM are too high compared to the GT at almost all sites, as shown in Fig.~\ref{fig:spatial_distribution}(A).
This is mainly because the $2500$ grid points extracted from the corresponding single low-resolution pixel tend to have similar values. 
In contrast, the results of the SRGAN family exhibit much sharper spatial contrast, e.g., the contrast between the north and south sides of the Chugoku area (around [36$^\circ$N, 135$^\circ$E]) is well captured.
{The differences in performance among these SRGAN-based methods are very subtle and a precise quantification is necessary to rank them: we will get back to this issue in the next paragraph.}
In Fig.~\ref{fig:spatial_distribution}(B,C), we qualitatively observe the same difference in the accuracy among the methods.
In particular, the SRGAN family, even including $\psi$SRGAN, successfully reproduce the nonmonotonic nature of the correlation as a function of the distance from the reference site: e.g., in the results of the GT and SRGAN-based methods in  Fig.~\ref{fig:spatial_distribution}(B), along the north side coastline (see the arrow in the figure), the correlation decays quickly near the reference point and then grows again around the Noto peninsula (around [38$^\circ$N, 137.5$^\circ$E]).
The CDFDM, on the other hand, merely exhibits the monotonic decay of the correlation along the same line.
{Please see also the SI Appendix for the difference plots between the GT and DS results.}

%------------------------------------------------------------------------------------------
%%% Table: Values of KL Divergence
%------------------------------------------------------------------------------------------
\tabcolsep = 5pt
\begingroup
\renewcommand{\arraystretch}{1.5}
\begin{table*}[t]
  \textbf{\caption{\label{table:MSE}Average MSE of the correlation coefficients}}
  \centering
  \begin{tabular}{c|cccc}
    \hline
     & $\pi$SRGAN & SRGAN  & $\psi$SRGAN & CDFDM \\
    \hline 
    $MSE_{\rm Jan}$ &${\bf 1.20\times 10^{-2}}$ &${1.27\times10^{-2}}$ &${2.04\times10^{-2}}$ &${4.35\times10^{-2}}$\\
    \hline
  \end{tabular}
\end{table*}
\endgroup

To quantify the accuracy of $C_{\rm Jan}^{DS}(l,l^\prime)$ for the different methods, we measure the mean square error (MSE) of the spatial distribution of the correlation coefficient defined as:
\begin{equation}
    MSE_M^{DS}(l)=\frac{1}{N_{\rm OS}}\sum_{l^\prime}^{N_{\rm OS}}(C_M^{DS}(l,l^\prime)-C_M^{\rm GT}(l,l^\prime))^2,
\end{equation}
where $l$ is the reference site and {$N_{OS}\equiv 630$} is the number of observation stations (see SI Appendix for a detailed explanation).
The values of {$MSE_{\rm Jan.}^{DS}$} measured based on each reference site are compared 
in Fig.~\ref{fig:spatial_distribution}(D), and the average values are listed in Table~\ref{table:MSE}} (the values for each site are shown in Table~S4 in the SI Appendix).
{All SRGAN family exhibit much better results than those of the CDFDM for all sites considered here and even $\psi$SRGAN offers twice better results.
Specifically, the best one, $\pi$SRGAN, achieves $3.6$ times better accuracy than the CDFDM for the average value over $12$ sites.
This result of the SRGAN-based methods being advantageous in achieving the ``naturalness'' of the spatial pattern is consistent with the report in ref.~\cite{PNAS_GAN_DS}.
If we further compare the results of SRGAN-based methods, although $\pi$SRGAN offers the best performance in terms of the mean value over all 12 sites, the standard SRGAN has the best values at the majority of locations, albeit by only small margins as shown in Figure 4D (and Table S4 in the SI Appendix).
We interpret this result as meaning that both $\pi$SRGAN and SRGAN demonstrate comparable performance in relation to the statistical characteristics of spatial correlation.
Together with the discussion in the previous subsections, the results presented in this section enable us to conclude that in the present $\pi$SRGAN, the auxiliary fields enhance the reproducibility of the simple statistics (such as $P(p)$) while maintaining the expression ability of the natural spatial expanse.
Such a strong downscaling ability highlights the applicability to local-scale and interregional assessments of climate change. }

\section*{Discussion}\label{sec:summary}
We have developed a machine learning-based statistical downscaling (DS) method with a large scale-factor of fifty, while maintaining both the basic statistical properties and the spatial correlation.
We employed a physics-informed type approach~\cite{Onishi2019} on the basis of the SRGAN-based method, and specifically, we developed a framework to use the proper auxiliary physical information along with the low-resolution input to attain large improvements in the DS performance
as summarized in Fig.~\ref{fig:this_method} and Tables~\ref{table:summary}, \ref{table:MSE}.
High accuracy comparable to the CDFDM, a conventional method in actual use, was demonstrated by directly comparing the climatological statistical properties with the real data.
More importantly, our approach exhibited the highly accurate reconstruction given in Fig.~\ref{fig:spatial_distribution} of the natural spatial distribution of the precipitation correlation coefficient, which was a serious issue for the conventional statistical DS methods, including CDFDM~\cite{SpatialCorrelation1,SpatialCorrelation2,SpatialCorrelation3}.
Since the importance of the multiregional spatial correlation has recently been recognized~\cite{MR1,MR2,MR3,MR4,MR5},
the present method is a promising new-generation alternative to conventional statistical DS methods, particularly in situations where the integration of the multiregional information is necessary. 

The detection and prediction of rare events are vital issues {\it inter alia} in the context of climate change assessments.
The methods including the present $\pi$SRGAN indeed have yet to accurately capture the low probability but significant rainfalls, as shown in Fig.~\ref{fig:test_precipitation_area}.
Here, we discuss possible directions to ameliorate the problem.
First, we could raise the level of physics-informed machine learning in terms of the classification proposed in ref.~\cite{Onishi2019}. 
If we succeeded in directly incorporating some part of the  governing equations into the learning process while maintaining the computational efficiency, local phenomena such as heavy rains would be predicted with high reliability.
Another direction is to take measures to reform the basic machine learning architecture itself.
Following the GAN-based approach, flow-based and diffusion model-based methods have attracted public attentions as powerful next-generation tools for general super-resolution tasks~\cite{SRFlow,SRDiff}. 
The main feature of these approaches is to generate multiple image candidates from a single input. 
Therefore, probabilistic information is expected to be drawn from the multiple super-resolved images, which would enable us to tackle the rare event predictions.

Another perspective concerns the use of machine learning techniques to improve the efficiency of dynamical downscaling, i.e., developing a high-speed machine-learning-based solver for the governing equations of climate models. 
Here we refer to an example of a speed up of multiscale simulations; in ref.~\cite{John} the Gaussian process is used to reduce the computational burden of multiscale simulation for polymeric liquid to achieve a reduction by a factor of $30$-$100$ without loss of accuracy.
Breakthroughs driven by similar approaches are expected once the complexity of the governing equations for the climate models is overcome.

{
Finally, we refer to the generalization ability of SRGAN.
Here, we have selected SRGAN instead of $\pi$SRGAN due to the anticipated lack of high generalization ability of the latter ($\pi$SRGAN relies on topographic information that is specific to the training area).
In the SI Appendix, we present the results of the generalization test, in which we tried to execute downscaling computations for samples derived from a different area than the one employed for training. 
Specifically, the test area encompasses the region spanning from 135.625 to 145.625 degrees east and from 35.625 to 45.625 degrees north, with a five-degree shift in both the eastward and northward directions from the original region used for the training. 
{The findings of the examination demonstrate considerably inferior performance compared to those reported in the main text, exposing the deficient generalization capability.
This suboptimal performance of the generalization ability is a somewhat predictable attribute since the training data are all from a specific same region. 
Even though we did not explicitly provide information about the topography in SRGAN, it is plausible that the network learned it indirectly through the temperature field, which exhibits a strong correlation with topography.}
We stress that we observe large errors even for Niigata and Kanazawa, which were part of the original computational domain.
To enhance the generalization ability, we would need to incorporate samples from a more extensive range of areas. 
The exploration of such an approach is left for future research.}

\section*{Methods}\label{sec:matmethods}
\subsection*{{CDFDM}}\label{sec:conventional_method}
Among a variety of statistical methods, we use, as a reference, the cumulative distribution function-based downscaling method (CDFDM) with quantile mapping that is in actual use.

If we simply map the low-resolution GCM simulation results onto the point at which the observations are available, we generally see a systematic difference, defined as bias, which comes from the systematic error of the model prediction and/or from the interpolation error. Removing this inherent bias is especially important in applying the downscaling results to the impact assessments.
In the CDFDM, bias is corrected via an empirical transfer function constructed in advance using measured data of distributional variables and the corresponding simulation results. 
The detailed procedure of constructing the transfer function is described as follows~\cite{Iizumi2010}.

The crude low-resolution data obtained from the GCM are first mapped onto a $2$\,km mesh using simple bilinear interpolation.
At each mesh point, an empirical cumulative distribution function (CDF) is then constructed using the interpolated data of the variable of interest over a specified time window. The transfer function is defined as a map of a variable onto the one at which the corresponding CDF of the observation falls within the same quantile level.
This preconstructed transfer function is applied under the assumption that the  error-percentile relation is conserved over time.
In the present study, the time window of a month is employed, while the original time window is over a half-year~\cite{Iizumi2010}, to more sensitively capture the seasonal trend~\cite{Yokohata2021,Ishizaki2022b}. 

Note that while this CDFDM is a nonparametric method, the corrected CDF perfectly matches the corresponding CDF of the observation (for the training data); the statistical properties of the downscaling results are expected to reproduce the observation well. 
The bias-corrected climate scenario obtained with this method has been widely used in climate change impact studies~\cite{Yokohata2021,Hiruta2022,Hiruta2022b}.

\subsection*{{SRGAN}}
We employ a generative adversarial networks-based (GAN-based) method as the basic machine learning architecture, which is called Super-Resolution Generative Adversarial networks (SRGAN) ~\cite{SRGAN}.
The terminology super-resolution (SR; or, in particular, single-image super-resolution) refers to a method of restoring a high-resolution image from the corresponding low-resolution data and is the counterpart of the downscaling in the realm of the general image processing.
The GAN-based methods are capable of generating realistic images by pitting a discriminator network against a generator network that generates samples (see Fig.~\ref{fig:this_method}A).
The discriminator network takes the real data (ground truths) and the fake data (output of the generator network) as inputs and identifies the authenticity of the input samples.
The generator network tries to deceive the discriminator while the discriminator tries to judge with high accuracy.
As a result, both networks spontaneously learn the ``realistic'' information.
The SRGAN can reproduce fine textures that cannot be achieved by normal convolutional neural network-based variants and offers substantially improved realistic super-resolution images.

Such network-based super-resolution techniques have recently been used for the DS tasks of climatological data. In a representative report by Stengel and coworkers, ref.~\cite{PNAS_GAN_DS}, the authors compared the performances of SRGAN-based downscaling methods with previous methods (SRCNN: Super-Resolution Convolutional Neural-Networks). 
Although the SRCNN-based method appeared to be superior in evaluating the performance in terms of the simple pixelwise MSE, the SRGAN-based method provided \emph{realistic} results satisfying the important physical requirements, e.g., the energy spectrum of the wind velocity field satisfied the Kolmogorov $5/3$ scaling law~\cite{Frisch1995} with remarkable accuracy.
The network architecture in our $\pi$SRGAN is mostly the same as the original SRGAN introduced in ref.~\cite{SRGAN}, although the batch normalization layers are removed obeying ref.~\cite{PNAS_GAN_DS}: {the explanation of the precise architecture is presented in SI Appendix}.
We also summarize other technical details, {such as the precise learning protocol, hyperparameter tuning, and the normalization of the data there.}
We note that the representative method compared to the $\pi$SRGAN referred to as ``SRGAN'' in our implementation is a slightly upgraded version including the high-resolution topography, which makes possible the decomposition of elements producing the improvement.
%\bibliography{sample}

\section*{Acknowledgements}

The authors thank N. Hanasaki and S. Koyama for fruitful discussions.
This research was partially supported by JST Grant Number JPMJPF2013.

\section*{Author contributions statement}

N.O. conducted numerical experiment, N.O., N.N.I., and H.Y. analyzed the data, N.O., S.K. and H.Y. invented the method, N.N.I., S.K., and H.Y. designed the work.  
All authors wrote the manuscript. 

\section*{Additional information}

\textbf{Competing interests} 
The authors declare no competing financial interests.

\noindent\textbf{Data availability} 
The datasets used and analysed in during this study are available from the corresponding author on reasonable request.

\bibliography{scibib}

\end{document}

\title{
Deep generative model super-resolves spatially correlated multiregional climate data--- supplemental material
}

\author{Norihiro Oyama}
\affiliation{Toyota Central R\&D Labs, Inc., Bunkyo-ku, Tokyo 112-0004, Japan}

\author{Noriko N. Ishizaki}
\affiliation{Center for Climate Change Adaptation, National Institute for Environmental Studies, Tsukuba 305-8506, Japan.}

\author{Satoshi Koide}
\affiliation{Toyota Central R\&D Labs, Inc., Bunkyo-ku, Tokyo 112-0004, Japan}

\author{Hiroaki Yoshida}
\affiliation{Toyota Central R\&D Labs, Inc., Bunkyo-ku, Tokyo 112-0004, Japan}

\maketitle

\section{Technical details}
\subsection{Architectures of networks}\label{sec:architecture}
\begin{figure*}[htb]
    \centering
    \includegraphics[width=\linewidth]{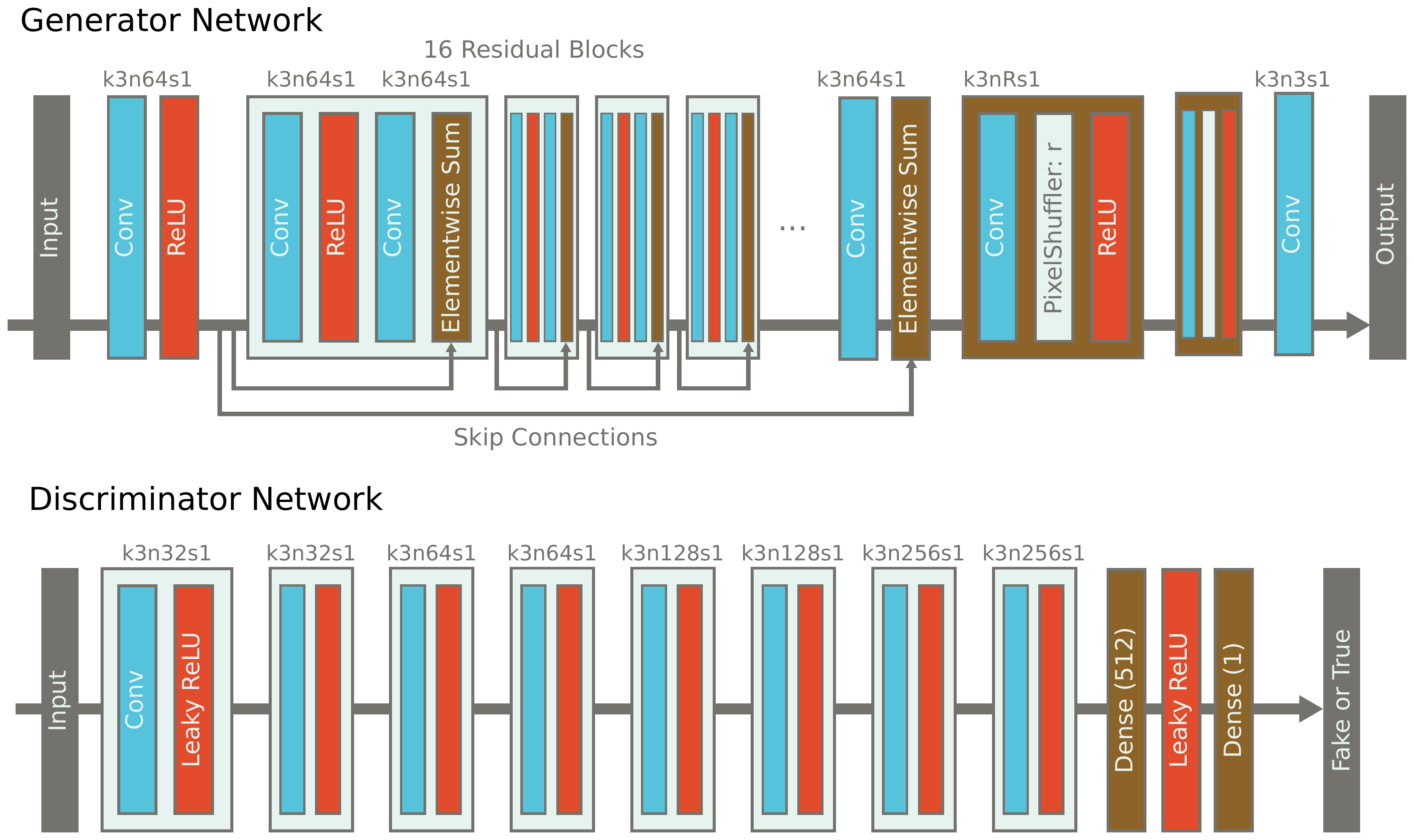}
    \caption{
    Schematic pictures of the architectures of the generator and discriminator networks.
}
    \label{fig:architecture}
\end{figure*}

The network architecture is depicted in Fig.~\ref{fig:architecture}.
In this figure, \emph{Conv} stands for the convolution layer, \emph{Dense} represents the fully connected layer, and \emph{ReLU} means Rectified Linear Unit.
The numbers and alphabets above the Convs represent the hyperparameters: the numbers after k, n, and s are the kernel size, the number of kernels, and the stride, respectively.
The number of kernels in the pixel shuffle layers $R$ is determined by the scale factor $r$ as $R=64\times r^2$.
There can be multiple pixel shuffle layers if the scale factor can be factorizable (in that case, each pixel shuffle layer has a factorized prime number value as the scale factor, e.g., 2 and 5 if the scale factor is 10). 
All Leaky ReLU layers in the discriminator network employ the slope of $0.2$ for negative inputs.

{
Following the conventional architectural design used in SRGAN for image processing, our generator network comprises of a pre-processing stage consisting of a convolution operation with ReLU activation, followed by 16 residual blocks equipped with skip connections, and pixel-shuffling units, culminating with a final convolution layer.
It is widely acknowledged that the use of residual blocks enables us to increase the depth of our network without encountering issues such as gradient loss, gradient explosion, or degradation. 
Additionally, the implementation of pixel-shuffling units is preferred over simple upsampling techniques like deconvolution due to their ability to eliminate undesired checkerboard pattern artifacts.
}

%Adding to the standard architecture presented above, we also introduced special processing in the input layer of the generator network to concatenate the high-resolution topographic information with other low-resolution information.
%This special treatment will be explained in Sec.~\ref{sec:input_process} later.

%%%%%%%

\subsection{Learning protocol}
In this section, we explain the details of the learning protocol.
The protocol basically obeys ref.~\cite{PNAS_GAN_DS}.
The fifty-fold downscaling is composed of two stages: the low-resolution to medium-resolution (LR-to-MR) and the medium-resolution to high-resolution (MR-to-HR) downscaling stages.
The scale factors for these two stages are ten- and fivefold, respectively, and a fifty-fold scaling factor is achieved overall.
The MR ground truths for the learning are generated by coarse-graining the HR image (average pooling with the kernel of size 5 and the stride is set to 5).
We also introduced pretraining for the Generator network as usual.
In total, there are four stages of learning: pretraining and GAN training for both LR-to-MR and MR-to-HR downscaling. 
We summarize the number of epochs and the batch size for each training stage in Table~\ref{table:batch_and_epoch}.
The same value of the initial learning rate, $r_l=10^{-4}$, is employed for all learning stages, and
the learning rate is multiplied by $0.99$ every epoch.
Under this exponentially decreasing-learning-rate protocol, the learning rate becomes approximately one tenth every 230 steps.
{Comparing the loss values for training and validation data during the training process, we confirmed that these combinations of the epoch numbers and the learning rate prevent the system from overtraining.
Other hyperparameters, such as the ones for Adam algorithm, are set to standard values used in most studies.}

%------------------------------------------------------------------------------------------
\tabcolsep = 5pt
\begingroup
\renewcommand{\arraystretch}{1.5}
\begin{table}[htb]
  \caption{Epoch number and batch size for each training state}
  \centering\label{table:batch_and_epoch}
  \begin{tabular}{c|cc|cc}
    \hline
     &\multicolumn{2}{c|}{LR to MR} &\multicolumn{2}{c}{MR to HR}   \\
    \hline
    & pre-training & GAN & pre-training & GAN\\
    \hline
    \hline 
    Epoch number  &200 &500 &200 &100\\
    Batch size &100 &100 &50 &5\\
    \hline
  \end{tabular}
\end{table}
\endgroup

Regarding the loss function, again, we obeyed ref.~\cite{PNAS_GAN_DS}.
We employed the simple MSE loss ${\cal L}_{\rm MSE}$ for the pretraining regardless of the resolution stages.
For the loss of the Generator network in the GAN part ${\cal L}_{\rm G}$, we use the linear combination of ${\cal L}_{\rm MSE}$ and the adversarial loss ${\cal L}_{\rm Adv}$:
\begin{align}
    {\cal L}_{\rm G}={\cal L}_{\rm MSE}+\alpha{\cal L}_{\rm Adv},
\end{align}
where $\alpha$ is the weight of the adversarial contribution and we set it as $\alpha=0.001$ in this work.
The adversarial loss represents the (in)accuracy of the discriminator classification of the outputs of the Generator (into fake and true).
For ${\cal L}_{\rm Adv}$, ref.~\cite{PNAS_GAN_DS} provides the detailed definition.
It is considered that the information can be extracted most efficiently from the Discriminator when the discriminator loss ${\cal L}_{\rm D}$ is around 0.5.
In this work, to maintain the value of ${\cal L}_{\rm D}$ in the vicinity of 0.5, adaptive training is carried out.
In this special training protocol, the training for the discriminator is repeated if ${\cal L}_{\rm D}$ is larger than 0.6 and that for the generator recurs if ${\cal L}_{\rm D}$ is less than 0.45.
Although in many cases such adaptive loops are performed for each minibatch, in this work, the adaptive loops are over the whole dataset.
We have employed these precise protocols because we found them to be better than other options, after trials and errors.

%%%%%%%
\subsection{Normalization of the data}\label{sec:norm}
All the data used for the training are normalized so that most (not all: see below) resulting pixels are in the interval $[0,1]$.
This is simply done by applying the following formula:
\begin{align}
    \tilde{x}=(x-x_{\rm lb})/x_{\rm scale},
\end{align}
where $x$ denotes the variable of interest (one of the temperature, precipitation, sea-level pressure, or topography), $\tilde{x}$ is the normalized value, and $x_{\rm lb}$ and $x_{\rm scale}$ are the lower bound and the scale of the variable $x$, respectively.
The precise values of $x_{\rm lb}$ and $x_{\rm scale}$ for each variable are summarized in Table~\ref{table:normalize}.
Such normalization is known to {allow the network to handle variables with different physical dimensions (in this study, temperature, precipitation, pressure, and altitude) in a unified manner and }enhance training efficiency.
We stress that for precipitation, this normalization does not guarantee that all the resulting values are less than unity since rare events exceed the threshold value of 100 mm/day (the maximum value in the test samples is 912 mm/day).
We employed this normalization factor because this choice gave the best performance among the values that we investigated (1000, 100, and 10 [mm/day]).

%------------------------------------------------------------------------------------------
\tabcolsep = 5pt
\begingroup
\renewcommand{\arraystretch}{1.5}
\begin{table}[htb]
  \caption{The parameters used for the data normalization}
  \centering\label{table:normalize}
  \begin{tabular}{c|cc}
    \hline
     & $x_{\rm lb}$ & $x_{\rm scale}$ \\
    \hline 
    Temperature & -50K & 100K\\
    Precipitation & 0 mm/day & 100 mm/day\\
    Sea level pressure & 950 hPa & 1050 hPa\\
    Topography & 0 m & 4000 m\\
    \hline
  \end{tabular}
\end{table}
\endgroup

%%%%%%%
\subsection{Precise measurement protocol of KL divergence}
{Technically, $D_{\rm KL}$ tends to infinity when $P^{\rm GT}(x)\ne 0$ and $P^{SR}(x)=0$ holds for a value $x$ (or more technically speaking, for a bin including $x$), and vice versa.
Problematically, this usually happens in real situations because of the limited number of samples: the probabilities of finding rare events are regarded as zero when the number of samples is finite even though they should be small but nonzero in the "true" distribution that is expected to be obtained when the sample number becomes infinite.
This results in an undesired infinite value of $D_{\rm KL}$.
To avoid such a trivial artifact, we applied the Gaussian smoothing function $f(x)=\frac{1}{\sqrt{2\pi}w}\exp{\frac{-(x)^2}{2w^2}}$ as:
\begin{align}
    {P}(x)=\frac{1}{Z\Delta}\sum_i^Nf(x-x_i),
\end{align}
where $w^2$ is the variance of the Gaussian and determines the smoothing width, $Z$ is the normalizing factor that guarantees $\int P(x)dx=1$, $\Delta$ is the bin width, and $N$ is the number of data.
$P(x)$ gives the probability of finding a sample in the interval $[x-\frac{\Delta}{2},x+\frac{\Delta}{2}]$ and $x_i$ is the value of the sample $i$.
We fixed this hyperparameter $w^2$ and $\Delta$ to be $w^2=16$ and $\Delta=2$ to obtain the results presented in Fig.~3 and Table~2 in the main text and Table~\ref{table:allKLDandMSE}. 
We confirmed that a change in the value of $w$ by a factor of 4 does not change the qualitative results.
}

%%%%%%%
\subsection{Locations where the correlation coefficients are evaluated}
If we calculate the correlations between all the $400\times 400$ grid points, the calculation cost becomes very expensive.
Therefore, we extracted only the grid points where the observation stations of the Automated Meteorological Data Acquisition System reside.
There are $N_{\rm OS}=630$ stations within the system boundary of our study.

%%%%%%%
\subsection{Topographic information}
The precise topographic information about the locations considered in Fig.~3 is summarized in Table~\ref{table:location}.
The pixels within the region specified by the min/max of the latitude/longitude compose each site.
Since, in our case studies, a single pixel in the high-resolution data has a linear dimension of 0.025 degrees in terms of both latitude and longitude, the regions of sites in Table~\ref{table:location} are all composed of 100 pixels.

%------------------------------------------------------------------------------------------
\tabcolsep = 5pt
\begingroup
\renewcommand{\arraystretch}{1.5}
\begin{table}[htb]
  \caption{Precise information of the location of each site}
  \centering\label{table:location}
  \begin{tabular}{c|cc|cc}
    \hline
     &\multicolumn{2}{c|}{Latitude} &\multicolumn{2}{c}{Longitude}   \\
    \hline
    & min & max & min & max\\
    \hline
    ({\bf A}) Akita &39.475$^\circ$N &39.725$^\circ$N &140.1$^\circ$E &140.35$^\circ$E\\
    ({\bf B}) Tokyo &35.625$^\circ$N &35.875$^\circ$N &139.65$^\circ$E &139.9$^\circ$E\\
    ({\bf C}) Niigata &37.65$^\circ$N &37.9$^\circ$N &139.0$^\circ$E &139.25$^\circ$E\\
    ({\bf D}) Shizuoka &34.95$^\circ$N &35.2$^\circ$N &138.25$^\circ$E &138.5$^\circ$E\\
    ({\bf E}) Nagoya &34.8$^\circ$N &35.05$^\circ$N &136.875$^\circ$E &137.125$^\circ$E\\
    ({\bf F}) Kanazawa &36.4$^\circ$N &36.65$^\circ$N &136.625$^\circ$E &136.875$^\circ$E\\
    ({\bf G}) Osaka &34.375$^\circ$N &34.625$^\circ$N &135.425$^\circ$E &135.675$^\circ$E\\
    ({\bf H}) Takamatsu &34.075$^\circ$N &34.325$^\circ$N &134.0$^\circ$E &134.25$^\circ$E\\
    ({\bf I}) Kouchi &33.525$^\circ$N &33.775$^\circ$N &133.55$^\circ$E &133.8$^\circ$E\\
    ({\bf J}) Izumo &35.125$^\circ$N &35.375$^\circ$N &132.75$^\circ$E &133.0$^\circ$E\\
    ({\bf K}) Hiroshima &34.225$^\circ$N &34.475$^\circ$N &132.525$^\circ$E &132.775$^\circ$E\\
    ({\bf L}) Oita &32.975$^\circ$N &33.225$^\circ$N &131.6$^\circ$E &131.85$^\circ$E\\
    \hline
  \end{tabular}
\end{table}
\endgroup

%%%%%%%

\section{Further details of results}
\subsection{Precise values of statistical indicators}
We summarize the precise values of $D_{\rm KL}$ and $MSE_{\rm Jan}$ for each site in Table~\ref{table:allKLDandMSE}.

%------------------------------------------------------------------------------------------
\tabcolsep = 4pt
\begingroup
\renewcommand{\arraystretch}{1.5}
\begin{table}[htb]
  \caption{Values of KL divergence of PDFs and MSE of correlation coefficients }
  \centering\label{table:allKLDandMSE}
  \begin{tabular}{c|cccc|cccc}
    \hline
    &\multicolumn{4}{c|}{$D_{\rm KL}$} &\multicolumn{4}{c}{$MSE_{\rm Jan}$}   \\
    \hline
     & $\pi$SRGAN & SRGAN &$\psi$SRGAN  & CDFDM & $\pi$SRGAN & SRGAN &$\psi$SRGAN  & CDFDM\\
    \hline 
    Akita 
    &{$1.62\times 10^{-3}$} &{$3.62\times 10^{-3}$} &{$3.03\times 10^{-2}$} &{$\bf 9.08\times 10^{-4}$}
    &{$1.43 \times 10^{-2}$} &{$\bf 1.32\times 10^{-2}$} &{$1.94\times 10^{-2}$} &{$4.43\times 10^{-2}$}\\
    Tokyo
    &{$3.40\times 10^{-3}$} &{$5.06\times 10^{-3}$} &{$5.36\times 10^{-3}$} &{$\bf 1.47\times 10^{-3}$}
    &{$6.41\times 10^{-3}$} &{$\bf 5.75\times 10^{-3}$} &{$8.96\times 10^{-3}$} &{$3.62\times 10^{-2}$}\\
    Niigata
    &{$8.17\times 10^{-3}$} &{$6.32\times 10^{-3}$} &{$2.01\times 10^{-2}$} &{$\bf 1.35\times 10^{-3}$}
    &{$1.75\times 10^{-2}$} &{$\bf 1.51\times 10^{-2}$} &{$3.20\times 10^{-2}$} &{$7.26\times 10^{-2}$}\\
    Shizuoka
    &{$2.63\times 10^{-3}$} &{$5.43\times 10^{-3}$} &{$4.15\times 10^{-3}$} &{$\bf 1.27\times 10^{-3}$}
    &{$8.40\times 10^{-3}$} &{$\bf 7.06\times 10^{-3}$} &{$1.13\times 10^{-2}$} &{$3.18\times 10^{-2}$}\\
    Nagoya 
    &{$2.32\times 10^{-3}$} &{$4.28\times 10^{-3}$} &{$8.72\times 10^{-3}$} &{$\bf 1.65\times 10^{-3}$}
    &{$\bf 8.48\times 10^{-3}$} &{$1.11\times 10^{-2}$} &{$1.61\times 10^{-2}$} &{$4.07\times 10^{-2}$}\\
    Kanazawa 
    &{$5.24\times 10^{-3}$} &{$4.24\times 10^{-3}$} &{$2.19\times 10^{-2}$} &{$\bf 1.46\times 10^{-3}$}
    &{$\bf 2.67\times 10^{-2}$} &{$3.91\times 10^{-2}$} &{$4.35\times 10^{-2}$} &{$8.77\times 10^{-2}$}\\
    Osaka
    &{$2.06\times 10^{-3}$} &{$3.79\times 10^{-3}$} &{$1.68\times 10^{-2}$} &{$\bf 1.51\times 10^{-3}$}
    &{$1.39\times 10^{-2}$} &{$\bf 1.25\times 10^{-2}$} &{$1.61\times 10^{-2}$} &{$4.55\times 10^{-2}$}\\
    Takamatsu
    &{$2.12\times 10^{-3}$} &{$2.65\times 10^{-3}$} &{$1.63\times 10^{-2}$} &{$\bf 1.35\times 10^{-3}$}
    &{$9.11\times 10^{-2}$} &{$\bf 8.67\times 10^{-3}$} &{$1.91\times 10^{-2}$} &{$4.41\times 10^{-2}$}\\
    Kouchi
    &{$2.73\times 10^{-3}$} &{$8.29\times 10^{-3}$} &{$1.07\times 10^{-2}$} &{$\bf 1.19\times 10^{-3}$}
    &{$9.23\times 10^{-3}$} &{$\bf 5.95\times 10^{-3}$} &{$9.74\times 10^{-3}$} &{$2.62\times 10^{-2}$}\\
    Izumo
    &{$1.20\times 10^{-3}$} &{$2.36\times 10^{-3}$} &{$3.81\times 10^{-3}$} &{$\bf 9.62\times 10^{-4}$}
    &{$\bf 1.28\times 10^{-2}$} &{$1.41\times 10^{-2}$} &{$2.98\times 10^{-2}$} &{$3.94\times 10^{-2}$}\\
    Hiroshima
    &{$\bf 1.80\times 10^{-3}$} &{$3.31\times 10^{-3}$} &{$4.87\times 10^{-3}$} &{$2.01\times 10^{-3}$}
    &{$7.69\times 10^{-3}$} &{$\bf 7.59\times 10^{-3}$} &{$1.81\times 10^{-2}$} &{$2.35\times 10^{-2}$}\\
    Oita
    &{$3.43\times 10^{-3}$} &{$3.23\times 10^{-3}$} &{$1.09\times 10^{-2}$} &{$\bf 2.88\times 10^{-3}$}
    &{$\bf 1.01\times 10^{-2}$} &{$1.19\times 10^{-2}$} &{$2.08\times 10^{-2}$} &{$3.05\times 10^{-2}$}\\
    \hline
    Average
    &{$3.06\times 10^{-3}$} &{$4.38\times 10^{-3}$} &{$1.28\times 10^{-2}$} &{$\bf 1.50\times 10^{-3}$}
    &{$\bf 1.20\times 10^{-2}$} &{$1.27\times 10^{-2}$} &{$2.04\times 10^{-2}$} &{$4.35\times 10^{-2}$}\\
    \hline
    \multicolumn{9}{r}{(the best values are shown in bold letters for each row, each indicator)}
  \end{tabular}
\end{table}
\endgroup

{
\subsection{Standard statistical indicators for single-image super-resolution}
In this subsection, we present customary statistical measures that are widely employed in the domain of image processing.
The first indicator is the mean squared error (MSE) of the pixel-based results ${\rm MSE}_{\rm pixel}$ defined as:
\begin{align}
\Delta_k &\equiv \frac{1}{N_l}\sum_l (p_k^{\rm GT}(l)-p_k^{\rm DS}(l))^2\\
{\rm MSE}_{\rm pixel} &\equiv \frac{1}{N_{\rm test}}\sum_k^{N_{\rm test}} \Delta_k
\end{align}
where $\Delta_k$ is the mean squared error of the $k$th sample and $N_l$ is the total number of pixels that represent the locations on land.
The remaining variables adhere to the definitions introduced in the main text.
Following the discussion in the main text, we consider only the precipitation here.

Using this pixel-wise measure ${\rm MSE}_{\rm pixel}$, the peak signal-to-noise ratio (PSNR), which is usually used as a quantitative measure of the performance of single-image super-resolution tasks, can be defined as:
\begin{align}
{\rm PSNR}_{P} = 10\cdot \log_{10}\frac{P^2}{\rm MSE}_{\rm pixel}
\end{align}
where $P$ stands for the maximum signal intensity.
In the case of image processing tasks, $P$ is trivially determined by the possible maximum intensity, e.g., it should be 255 if the pixel is expressed by 8-bit information.
However, in the current situation, the peak signal intensity is not known in advance and thus we employed the maximum values observed in the test samples.
There are ``two different peak signals'': the one observed in the downscaling results and the one in the ground truth.
In Table~\ref{table:psnr}, we compare the values of PSNR that are measured using the peak signal of the ground truth $P_{\rm GT}=9.12$ and the ones in the downscaling results $P_{\rm DS}$.
Despite a very large scale factor of fiftyfold, the results for all methods are remarkably favorable when compared to the benchmarks established in the image processing field.
This is simply because the precipitation field exhibits very small values in most sites in most samples compared to the peak signal, and so are the errors.

%------------------------------------------------------------------------------------------
\tabcolsep = 5pt
\begingroup
\renewcommand{\arraystretch}{1.5}
\begin{table}[htb]
  \caption{MSE, PSNR, and $P_{\rm DS}$ of machine learning-based methods}
  \centering\label{table:psnr}
  \begin{tabular}{c||c|cc|c}
    \hline
     & ${\rm MSE}_{\rm pixel}$ & ${\rm PSNR}_{P_{\rm GT}}$ & ${\rm PSNR}_{P_{\rm DS}}$  &$P_{\rm DS}$\\
    \hline 
    $\pi$SRGAN & $7.99\times 10^{-3}$ & 40.2 & 39.1 & 8.10\\
    SRGAN & $8.35\times 10^{-3}$ & 40.0 & 37.6 & 6.90\\
    $\psi$SRGAN & $1.15\times 10^{-2}$ & 38.6 & 36.4 & 7.10\\
    \hline
  \end{tabular}
\end{table}
\endgroup
%%%%%%%

To offer a more intuitive understanding of the difference in the accuracy of each approach ($\pi$SRGAN, SRGAN, $\psi$SRGAN, and CDFDM), we have included Fig.~\ref{fig:diff1}, which illustrates the difference plots for the single-sample visualization (corresponding to Fig.~2 in the main text), and Fig.~\ref{fig:diff2}, which depicts the spatial distribution of the correlation coefficient (Fig.~4 in the main text).
These figures substantiate the assertions made in the main text.

In Fig.~\ref{fig:diff1}~(A), the disparities between the temperature fields of the ground truth and those generated by the downscaling techniques are illustrated. 
Interestingly, the magnitude of the error is minimal in CDFDM and maximal in $\psi$SRGAN in accordance with the statistical metrics of precipitation.
However, since $\psi$SRGAN incorporates the LR information about the temperature field like other SRGAN-based methods, this inferior performance for the temperature field is unexpected. 
We need to conduct a comprehensive statistical analysis to draw a definitive conclusion: such analysis falls outside the scope of this paper.

 Fig.~\ref{fig:diff1}~(B) shows the differences in the precipitation field between the ground truth and the downscaling results.
 Here, while large errors are relatively widely distributed in the cases of $\psi$SRGAN and CDFDM, only small and localized errors are seen in the cases of $\pi$SRGAN and SRGAN.

In Fig.~\ref{fig:diff2}, the errors in the spatial distribution of the correlation coefficient (again, defined as the simple root mean squared errors from the ground truth) are displayed. Notably, in the instances of CDFDM, we note extensive regions exhibiting dark colors, which denote significant discrepancies. 
Specifically, with respect to CDFDM, the areas of dark color align with prominent ridgelines. This is indicative of the spatial layout of the correlations, originating from the topography, being inadequately represented by CDFDM.
On the other hand, in the columns of SRGAN-based methods, again even including $\psi$SRGAN, we observe light hues across nearly all locations. 
This outcome serves to demonstrate the exceptional precision of SRGAN-based methods and, in particular, our approach $\pi$SRGAN.

\begin{figure*}[htb]
    \centering
    \includegraphics[width=\linewidth]{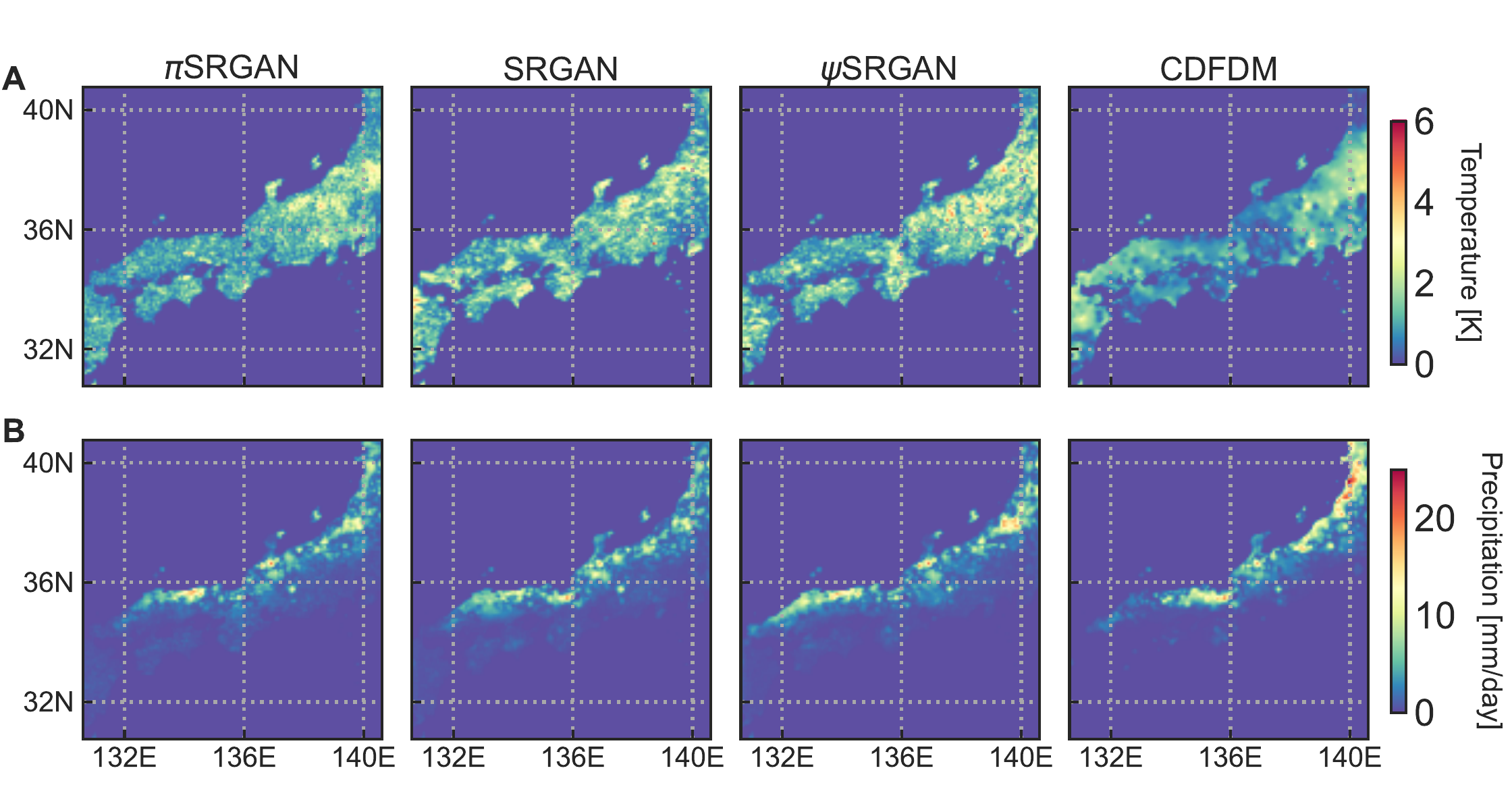}
    \caption{
    Difference plots for single-sample visualization.
    Distributions of errors in (A) temperature and (B) precipitation obtained from $\pi$SRGAN, SRGAN, and CDFDM are compared.
    The data from January 24, 2008, are displayed, as in Fig.~2 in the main text.
}
    \label{fig:diff1}
\end{figure*}

\begin{figure*}[htb]
    \centering
    \includegraphics[width=\linewidth]{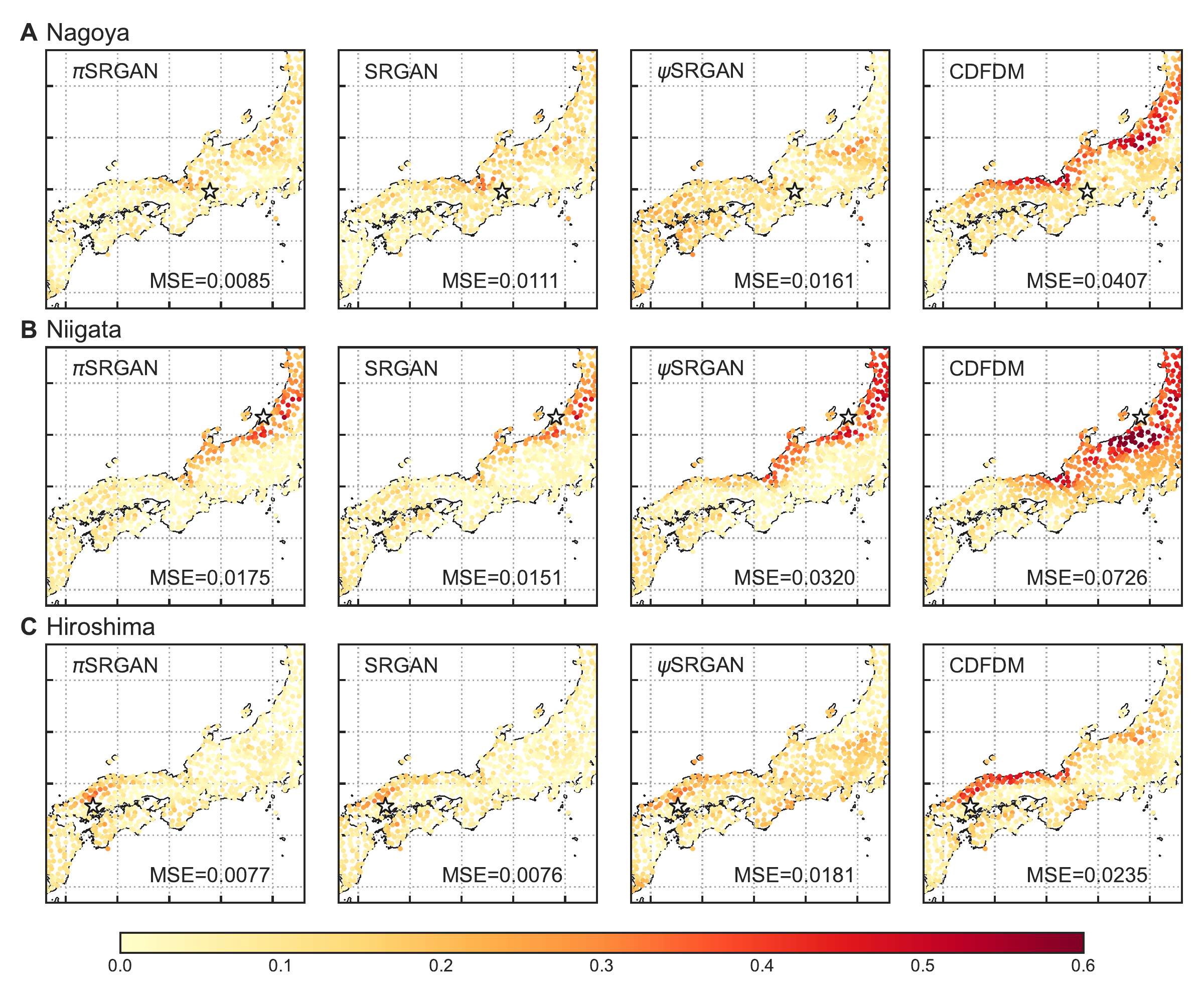}
    \caption{
    Spatial distribution of the mean squared error in the correlation coefficient for precipitation.
        The results of January obtained with the $\pi$SRGAN, SRGAN and CDFDM are compared in the case of the reference point of correlation at Nagoya {[35.1667$^\circ$N, 136.965$^\circ$E]} (A), Niigata {[37.9133$^\circ$N, 139.0483$^\circ$E]} (B), and Hiroshima {[34.365$^\circ$N, 132.4333$^\circ$E]} (C).
    The dot color indicates the values of $|C_{\rm Jan}^{GT}(l,l^\prime)-C_{\rm Jan}^{DS}(l,l^\prime)|$ between the location of the dots and the reference site.
    The reference points are represented by star symbols.
}
    \label{fig:diff2}
\end{figure*}

\section{Generalization ability test}
In this section, we present the outcomes of the generalization capability test of SRGAN.
Specifically, we performed downscaling computations on the test samples covering the region spanning from 135.625 to 145.625 degrees east and from 35.625 to 45.625 degrees north, utilizing the Generator network featured in the main text. 
The network was initially trained on the area illustrated in Fig.~3M in the main text, ranging from 130.625 to 140.625 degrees east and 30.625 to 40.625 degrees north. 
Thus, the samples employed in this generalization test were shifted by five degrees both in the north and east directions, relative to those utilized in the original analysis.
We have calculated the simple probability distribution function of the precipitation $P(p)$ and $D_{\rm KL}$ between $P^{\rm GT}(p)$ and $P^{\rm SRGAN}(p)$ (the information presented in Fig.~3) for 8 representative locations depicted in Fig.~\ref{fig:gen}~(I).
Notice that two out of these eight sites are the same ones considered in Fig.~3 in the main text (Niigata and Kanazawa).
The results of $P(p)$ are shown in Fig.~\ref{fig:gen}~(A-H) and $D_{\rm KL}$ for each site is shown in Fig.~\ref{fig:gen}(J).
We note that, unlike Fig.~3(P) in the main text, we needed to employ the logarithmic scale for the ordinate of Fig.~\ref{fig:gen}(J) because the accuracy becomes worse by orders of magnitude for several sites (e.g., Akita and Kanazawa).
These results indicate that the precision of our method deteriorates significantly when we alter the target location for the downscaling (quantitatively speaking, the value of the KL divergence becomes more than ten times higher).
 Errors are observed not only in the high precipitation regime, where $P^{\rm GT}(p)$ attains very small values, but also in the low precipitation regime, e.g., in Fig.~\ref{fig:gen}(B,E,G,H).
The absence of generalization capacity is an anticipated characteristic as our network has been optimized for a specific geographical region.

\begin{figure*}[htb]
    \centering
    \includegraphics[width=.8\linewidth]{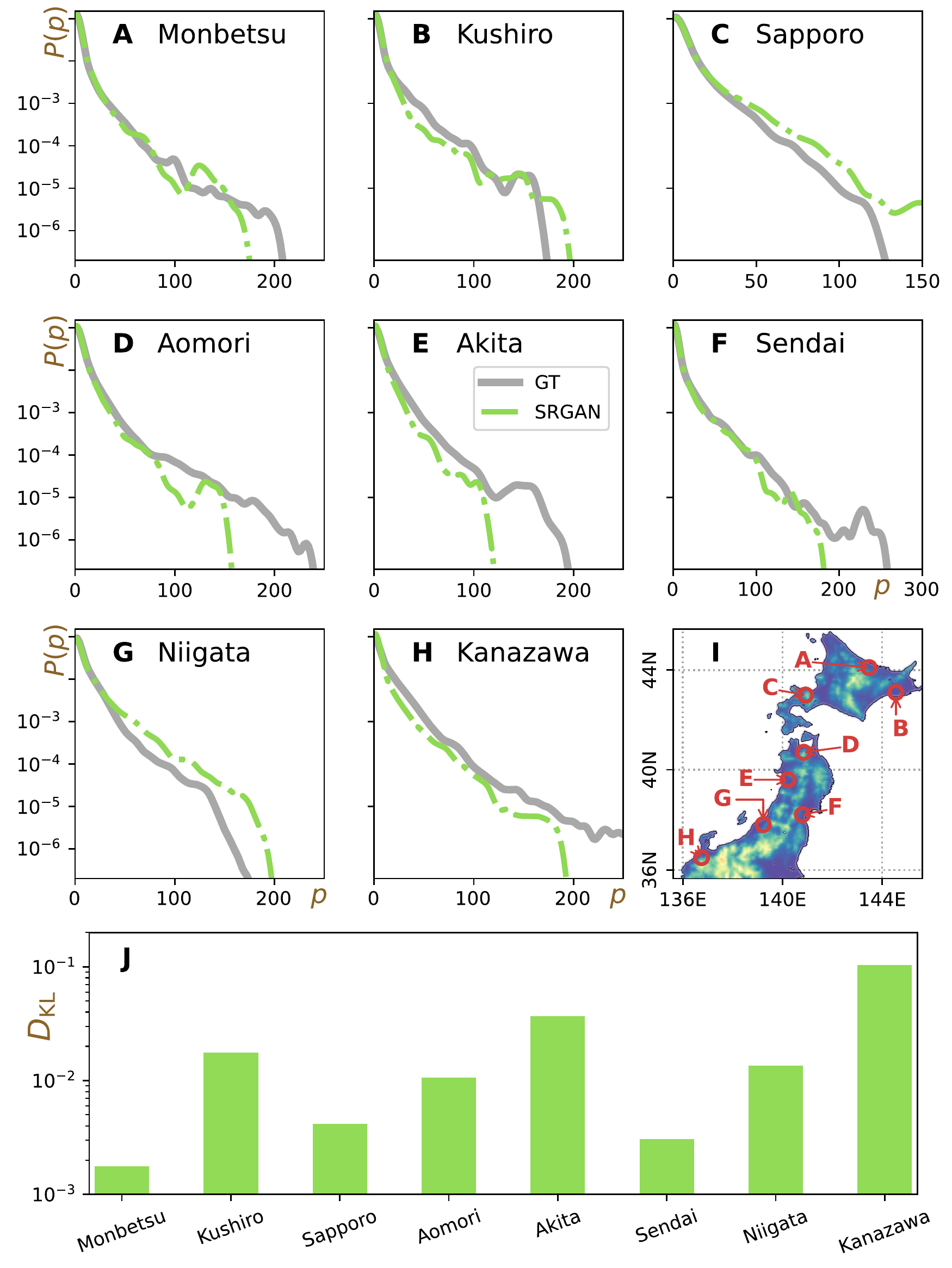}
    \caption{
    Results of generalization ability test: statistics of precipitation calculated for a region different from ones used for training.
   (A-H) The probability distribution functions (PDFs) $P(p)$ as a function of the precipitation $p$ at each site. 8 representative sites are chosen from the entire computational domain of the test.
   The different ranges of the abscissa are employed for different sites. 
   (I) The normalized topographic information and the locations of $8$ sites of panels (A-H).
   (J) Bar plot of the Kullback-Leibler divergence $D_{\rm KL}$.
}
    \label{fig:gen}
\end{figure*}
}
\bibliography{scibib}